\documentclass[useAMS,usenatbib]{mn2e}
\usepackage{latexsym,graphicx,natbib,amssymb,array,aas_macros}

\newcommand{\msun}{M_{\odot}}
\newcommand{\mmax}{m_{\rm max}}
\newcommand{\tM}{\tau_M}
\newcommand{\tcr}{t_{\rm cr}}
\newcommand{\mcl}{M_{\rm cl}}
\newcommand{\mecl}{M_{\rm ecl}}
\newcommand{\mgas}{M_{\rm gas}}
\newcommand{\mpl}{M_{\rm pl}}
\newcommand{\mpc}{M_{\odot}\;{\rm pc}^{-3}}
\newcommand{\rhoecl}{\rho_{\rm ecl}}
\newcommand{\rhocl}{\rho_{\rm cl}}
\newcommand{\rh}{r_h}
\newcommand{\rpl}{r_{\rm pl}}
\newcommand{\rhi}{r_{h,i}}
\newcommand{\rhf}{r_{h,f}}
\newcommand{\adrei}{\alpha_3}

\newcommand{\Ein}{E_{\rm in}}
\newcommand{\Efin}{E_{\rm fin}}
\newcommand{\Ereq}{E_{\rm OB}^{\rm req}}
\newcommand{\Epl}{E_{\rm OB}^{\rm req}}
\newcommand{\Edot}{\dot E}
\newcommand{\EOB}{E_{\rm OB}^{\tM}}
\newcommand{\ml}{M_{\rm dyn}/L}

\title[Top-heavy IMFs in GCs]{Evidence for top-heavy stellar initial mass functions with increasing density and decreasing metallicity}
      
\author[M. Marks, P. Kroupa, J. Dabringhausen and M. S. Pawlowski]
{
  Michael Marks$^{1,2,}$\thanks{Member of the International Max Planck Research School (IMPRS) for Astronomy and Astrophysics at the Universities of Bonn and Cologne; e-mail: mmarks@astro.uni-bonn.de (MM)}, Pavel Kroupa$^1$, J\"org Dabringhausen$^1$ and Marcel S. Pawlowski$^1$\\
  $^1$Argelander Institute for Astronomy, University of Bonn, Auf dem H\"ugel 71, 53121 Bonn, Germany\\
  $^2$Max-Planck-Institut f\"ur Radioastronomie, Auf dem H\"ugel 69, D-53121 Bonn, Germany\\
}       

\begin{document}

\date{Accepted ????. Received ?????; in original form ?????}

\pagerange{\pageref{firstpage}--\pageref{lastpage}} \pubyear{2010}

\maketitle

\label{firstpage}

\begin{abstract}
Residual-gas expulsion after cluster formation has recently been shown to leave an imprint in the low-mass present-day stellar mass function (PDMF) which allowed the estimation of birth conditions of some Galactic globular clusters (GCs) such as mass, radius and star formation efficiency. We show that in order to explain their characteristics (masses, radii, metallicity, PDMF) their stellar initial mass function (IMF) must have been top-heavy. It is found that the IMF is required to become more top-heavy the lower the cluster metallicity and the larger the pre-GC cloud-core density are. The deduced trends are in qualitative agreement with theoretical expectation. The results are consistent with estimates of the shape of the high-mass end of the IMF in the Arches cluster, Westerlund~1, R136 and NGC~3603, as well as with the IMF independently constrained for ultra-compact dwarf galaxies (UCDs). The latter suggests that GCs and UCDs might have formed along the same channel or that UCDs formed via mergers of GCs. A fundamental plane is found which describes the variation of the IMF with density and metallicity of the pre-GC cloud-cores. The implications for the evolution of galaxies and chemical enrichment over cosmological times are expected to be major.
\end{abstract}


\section{Introduction} \label{sec:intro}
\subsection{Are PDMF variations evidence for IMF variations?}
The stellar present-day mass function (PDMF) is observed to be different for individual globular clusters (GCs) of the Galaxy. Their low-mass stellar content differs significantly \citep{mcc86,dpc93,dmpp07,Paust2010}.

These differences are claimed to be explainable by a universal initial mass function (IMF) and secular two-body relaxation driven evolution alone \citep{Leigh2012}, in agreement with earlier work that shows that an initially rising IMF with decreasing stellar mass can bend over in the course of a Hubble time \citep{vh97,bm03}. \citet{dmpp07} however noticed that clusters which are strongly depleted in low-mass stars have a low concentration while no such cluster with a high concentration exists. This trend has been argued to be incompatible with standard secular evolution by \citet{dmpp07} unless the majority of clusters with a flat or even declining low-mass PDMF are post-core collapse clusters. \citet{Paust2010} question the existence of such a trend since no very strongly depleted cluster is present in their sample. Their data however still closely follows the relation proposed by \citet{dmpp07}.

Primordial mass-segregation appears to come to the rescue of the secular evolution picture, since if low-mass stars form preferentially at the cluster outskirts they are easily stripped off the cluster by the tidal-field of the host galaxy. There is then no need for the cluster to go into core-collapse which would otherwise be necessary to drive the preferential evaporation of low-mass stars \citep{bdmk08}. However, these authors also show that some clusters reported by \citet{dmpp07} which are most strongly depleted in low-mass stars, can only be understood if clusters fill their tidal-radii. Tidally underfilling clusters can not reach the observed degree of low-mass star depletion, even when they are close to dissolution \citep{bdmk08}. Young star clusters of the Galaxy are however observed to be very compact \citep{ll03}, not closely filling their tidal limits and such initial conditions cannot be the dominant cause of low-mass star depletion.

Is the concentration--PDMF trend observed by \citet{dmpp07} therefore the first-time evidence for a variable low-mass IMF?  \citet{mk10} show that clusters with a flat or declining PDMF have the largest metallicity in the \citet{dmpp07} sample of clusters. Such a trend is difficult to understand in the dynamical evolution picture since it is unclear how dynamics could possibly know about the metal-content of the cluster. The trend is also difficult to understand if the low-mass IMF were to vary since relatively fewer low-mass stars would have to form with decreasing metallicity, in contradiction to theoretical expectation \citep{Bastian2010,KroupaRev2012}. Furthermore the concentration--PDMF trend is also difficult to reconcile with the assumption of a varying low-mass IMF: In dense, i.e. highly concentrated clusters, low-mass stars might disappear through merging in dense clusters so that low-mass stars would be expected to be underabundant in highly concentrated clusters, which is again contrary to the \citet{dmpp07} observation.

Thus, if standard two-body relaxation alone is not sufficient in removing low-mass stars and a varying low-mass IMF appears not to be a feasible solution, where do the observed differences in the PDMF come from? \citet{mkb08a} propose a residual-gas expulsion scenario which adresses all these issues at once. They show that quick gas removal from compact, primordially mass-segregated clusters with binaries leads to low-concentration clusters which are depleted in low-mass stars, while initially binary-rich clusters with slow gas removal retain their input IMF, as observed by \citet{dmpp07}. The expansion following gas expulsion (accompanied by the loss of primarily low-mass stars) naturally leads to tidally-filling clusters with low concentration. This process is expected to be metallicity dependent since metal-rich material will couple better to the radiation driving the removal of the residual-gas \citep{mk10}. The resulting quicker gas expulsion, in turn, leads to more metal-rich clusters being more strongly depleted of low-mass stars, as observed. This gas expulsion scenario is in qualitative agreement with the finding by \citet{Strader2009,Strader2011} that mass-to-light ratios for M31 GCs are lower in relatively more metal-rich environments.

Thus, invoking gas expulsion there is no need for a variation in the low-mass IMF. In contrast, for stars more massive than $\approx1\msun$ theoretical arguments suggest that the massive star content should depend on the ambient conditions.

The Jeans-mass ($M_J\propto\rho^{-1/2}T^{3/2}$) is higher for denser \citep{l98,BateBonnell2005,Bonnell2006} and warmer gas \citep{Klessen2007} such that a higher average mass for stars is expected when favourable conditions are met. As both the temperature and density of the gas depend on its metallicity through less efficient cooling in lower metallicity environments, more massive stars should form from low metallicity gas. Additionally, forming stars self-regulate their masses via radiative feedback \citep{AdFat96}. Feedback is also expected to be metallicity dependent, as photons couple less efficiently to gas of lower metallicity. Anew, a lower metallicity should favour the formation of more-massive stars. Therefore both the Jeans-mass-- as well as the self-regulation--arguments suggest the fraction of high to low mass stars to increase with decreasing metallicity.

Also, star clusters have been found to be rather compact when they reach their densest state \citep{Testi98,Kroupa05,Scheep09,mk10}. The resulting density of stars is expected to have an influence on the IMF \citep{Bonnell98,Shad04}. \citet{ElmShad03} concluded that densely packed stars should lead to \emph{top-heavy} IMFs, i.e. an IMF with more massive stars than expected from the canonical \citet{k01} IMF (equation~\ref{eq:imf} below), in the most massive clusters (e.g. GC progenitors) through merging. \citet{Weid10} show, using a geometrical argument, that a high density will be a problem for the formation of individual stars in clusters more massive than $\approx10^5\msun$. And, cluster formation at high star formation rates may lead to top-heavy IMFs by the heating of molecular gas through supernovae generated cosmic rays \citep[the Papadopoulos CR-heating,][]{Pap10}.

Finding observational evidence for a varying high-mass IMF is difficult. In GCs, stars more massive than $1\msun$ have long since evolved away from the main-sequence and cannot be observed. Rather weak direct observational evidence for top-heavy IMFs in young starburst clusters exists:

The Arches-cluster ($2-4$~Myr), only $\approx25$~pc in projected distance from the Galactic centre, should be subject to a very strong tidal-field and is a prime candidate which has been reported to host a non-canonical population of high-mass stars. The measurements however differ widely \citep[$\adrei=1.65-2.3$ in equation~\ref{eq:imf},][]{Figer1999,Stolte2002,Kim2006,Espinoza2009}. The same holds true for the NGC 3603 cluster ($2-3$~Myr), about $8$~kpc from the Galactic centre and $6$~kpc from the Sun. Its reported MF slope ranges from $\adrei=1.9$ to $2.3$ \citep{Nuerberger2002,Sung2004,Stolte2006,Harayama2008}. The most massive known young cluster in the Galaxy Westerlund 1 (Wd1, $4-7$~Myr) in contrast has $\adrei=2.3$ between $0.75$ and $1.5$~pc and $\adrei=1.6$ closer in \citep{Brandner2008}. The 30~Dor region ($\approx3$~Myr) in the Large Magellanic Cloud (LMC), having a mass comparable to Wd1, also exhibits no deviation from the canonical IMF \citep[excluding the central cluster R136]{SelmanMelnick2005}. R136 has $\adrei=2.2-2.6$ \citep{Brandl1996,Massey1998,Andersen2009}. All reported clusters however show evidence for mass segregation, i.e. the results might be biased through dynamical effects \citep[see, e.g.,][]{pz07}.

Better evidence for top-heavy IMFs is seen in ultra-compact dwarf galaxies (UCDs). These are objects similar to GCs, given the ideas of their formation from star cluster complexes \citep{Fel05} or as the most massive GCs \citep{mieske02} and are thus ideally suited to be compared to GCs. Observed dynamical-mass-to-light ($\ml$)-ratios appear too large when they are compared to expectations from canonical stellar population models with the canonical IMF \citep{dab08}. \citet{dab10} explain these observations by invoking top-heavy IMFs for them, i.e. UCDs had non-canonical IMFs above $1\msun$. The subsequent, primarily gas expulsion driven evolution then turns them into objects which resemble the observed properties of UCDs today. The \citet{dab10} simulations show that top-heavy IMFs with $1\lesssim\adrei<2.3$ for proto-UCDs are indeed viable to explain the presently observed $\ml$-ratios in them and their models resemble present-day UCDs in mass and size. This result is significantly strengthened by the overabundance of X-ray-bright UCDs which is in excellent agreement with the surplus of neutron-stars and black holes expected for a top-heavy IMF \citep{Dab2011}.

In this contribution it is shown that evidence for a top-heavy IMF also emerges for GCs if residual-gas expulsion from mass-segregated clusters starting with a universal low-mass IMF is responsible for the observed low-mass star depleted PDMFs \citep{dmpp07}, as suggested by \citet{mkb08a}. In order to deplete a universal low-mass IMF in GCs to the observed degree through the gas throw-out phase the IMF must have been top-heavy in order for the high mass stars to be sufficiently destructive to the gas content of the forming GCs.

\subsection{Tracing the high-mass IMF in GCs via residual-gas expulsion} \label{sec:trace}
The IMF at the high-mass end can never be directly seen in Galactic GCs, since massive stars are short-lived while GCs are old objects. Stars above the turn-off mass for GCs ($\approx0.8\msun$) cannot be observed. However, the distribution of high-mass stars, which were present in a GC after its birth, can possibly be traced by indirect means.

Young gas-embedded star clusters expell their left-over gas from star formation due to stellar feedback over a time-scale $\tM$ \citep{kah01}. For a given mass, size and star formation efficiency of a GC's pre-gas expulsion cloud-core (containing stars+gas), $\tM$ is determined by the number of massive O- and B-stars, which are the main donators of energy into the gas and, in turn, drive the gas expulsion process. Thus, the shape of the high-mass end of the IMF of stars defines $\tM$. For a star formation efficiency (SFE),
\begin{equation}
 \epsilon=\frac{\mecl}{\mecl+\mgas}\;,
\end{equation}
$0.1<\epsilon<0.5$, rapid gas expulsion (short~$\tM$) gives rise to a phase of strong expansion following gas expulsion and the cluster will loose stars over the tidal boundary \citep{bk07}. If a cluster is mass-segregated at birth, as indicated by theory \citep{mkb08a,bdmk08,Allison09} and observations \citep{Little03,McCrad05,Chen07}, the cluster will preferentially loose its low-mass stars. \emph{The form of the high-mass IMF can thus leave an imprint on the low-mass PDMF via the gas expulsion process}.

Using the \citet{dmpp07} concentration, $c=\log(r_t/\rh)$, versus low-mass PDMF slope diagram as a diagnostic tool, SFEs, $\epsilon$, pre-GC cloud-core masses, $\mcl$, half-mass radii, $\rh$, and densities, $\rhocl$, at star cluster birth are constrained for the sample of $20$ GCs in \citet{dmpp07} for which PDMFs have been measured down to $0.3\msun$ by comparison of observations with $N$-body models \citep{mk10}. These cluster parameters will in the following be referred to as the \emph{initial conditions} of GCs and are reported in Table~\ref{tab:tab}.

In order to remove the gas from these clusters completely it needs to travel between $\approx0.5-1.5$~pc from the cluster centre to leave the cluster (compare to the $\rh$ reported in Table~\ref{tab:tab}). If the gas travels with the speed of sound in the ionised ISM (about $10$~km~s$^{-1}\approx10$~pc~Myr$^{-1}$) the gas expulsion time, $\tM$, becomes $0.5-1.5\times10^5$~yr. This is an estimate for the shortest timescale on which gas can be expelled from a star cluster. It will be shown that gas cannot be removed within $\tM\approx10^5$~yr ($\equiv\tM$ for a cluster of size $1$~pc) if the IMF above $1\msun$ has been canonical but that instead a top-heavy IMF in dependence of the environment needs to be invoked ($\adrei$ in equation~\ref{eq:imf} varies).

\subsection{The models}
The following analysis relies on the \citet{mk10} results. In order to obtain the initial conditions they assumed residual-gas expulsion to be the dominant phase driving the low-mass star depletion in GCs and to be the source of the \citet{dmpp07} concentration--PDMF slope relation. They used the residual-gas expulsion $N$-body model grid by \citet{bk07} to infer the mass loss and expansion through gas removal from embedded clusters. \citet{mk10} assume that post-gas expulsion cluster expansion is only weak compared to expansion driven by gas expulsion. \citet{bdmk08} calculated initial masses for the \citet{dmpp07} sample of GCs assuming two-body relaxation driven low-mass star depletion without gas, which \citet{mk10} used as cluster masses after residual-gas throw out in order to account for mass loss through long-term secular evolution. This finally allowed to trace-back present-day cluster masses and sizes to their initial values.

While the \citet{mk10} results depend on the validity of these assumptions, it is the first time that initial conditions have been constrained for a large sample of real clusters. Undoubtedly improved initial conditions could be obtained by running $N$-body models including gas expulsion and two-body relaxation driven evolution afterwards to self-consistently compare observational data with computations. Such $N$-body models of initially compact $\mecl>10^5\msun$ clusters are currently however prohibitively expensive computationally. For the time being their results constitute the best available homogeneously obtainted constraints on GC initial parameters. The results derived here should therefore not be considered to be a very final word but they allow at the very least an important insight into star formation in massive star-bursting clusters at high redshift.

\section{Constraining $\adrei$} \label{sec:tM}
Observationally inferred stellar initial mass functions (IMFs) of stars are found to be indistinguishable from the \emph{canonical} form, $\xi_c(m)$. Stellar IMFs are conveniently represented by a multi-part power-law that describes the number of stars, $dN$, forming in the mass interval $[m,m+dm]$,
\begin{equation}
 dN/dm=\xi(m)=k\;a_i\times m^{-\alpha_i}\quad(i=1,2,3)\;,
 \label{eq:imf}
\end{equation}
where $k$ is a normalization constant and the $a_i$ warrant continuity at the edges of the power-law segments. For main-sequence stars the canonical IMF has the slopes $\alpha_{1,\rm c}=1.3$ for $m/\msun\in$ [$0.08,0.5$] and $\alpha_{2,\rm c}=\alpha_{3,\rm c}=2.3$ for $m/\msun\in$ [$0.5,1$] and [$1,\mmax$], respectively \citep{k01}. The physical stellar upper-mass limit is $\mmax\lesssim150\msun$, which depends on the mass of the cluster \citep{Weid10}.

Assuming the IMF to be canonical below $1\msun$, we seek to infer the slope, $\adrei$, in order to remove the residual-gas within a given time-scale, $\tM$.

The idea is to compare the energy required to remove the gas, $\Ereq$, with the energy, $\EOB$, provided by O- and B-type stars within $\tM$. $\Ereq$ is the difference between the initial binding energy of the cluster (with gas) and the binding energy after the gas has been expelled,
\begin{equation}
 \Ereq=\Ein-\Efin\;.
 \label{eq:ediff}
\end{equation}
For a mass, $\mpl$, distributed according to a Plummer density profile, the potential energy is
\begin{equation}
 \Epl=\frac{3\pi}{32}\frac{G\mpl^2}{\rpl}\;,
\end{equation}
where $\rh=1.305\;\rpl$ is the characteristic Plummer radius \citep{Kroupa2008}. Then equation~(\ref{eq:ediff}) becomes
\begin{eqnarray}
  \Ereq&=&1.305\times\frac{3\pi\;G}{32}\left(\frac{\mcl^2}{\rhi}-\frac{\mecl^2}{\rhf}\right) \nonumber \\
  &=&1.305\times\frac{3\pi\;G}{32}\left(\frac{\mcl^2}{\rhi}-\frac{\mecl}{\mcl}\frac{\mecl^2}{\rhi}\right)\;,
  \label{eq:ereq}
\end{eqnarray}
where $\mecl=\epsilon\mcl$ is the mass in stars, $\rhi\equiv\rh$ is the initial half-mass radius \citep[both from][]{mk10} and $\rhf$ is the final half-mass radius of the cluster after the gas has been expelled. The last equality holds if the gas is removed adiabatically, i.e. slow with respect to the clusters crossing-time since then \citep{h80}
\begin{equation}
 \rhf=\rhi\;\frac{\mcl}{\mecl}\;.
 \label{eq:hills}
\end{equation}
The crossing-times,
\begin{equation}
 \tcr=\frac{2}{\sqrt{G}}\mcl^{-1/2}\rh^{3/2}\;,
\end{equation}
for the investigated sample of GCs as calculated from the initial conditions derived in \citet{mk10} are indeed much shorter than the here assumed residual-gas expulsion times.

The rate by which radiative plus mechanical energy from all stars is deposited in the interstellar medium (ISM) is,
\begin{equation}
 \Edot=\int_{0.08\msun}^{\mmax}\dot{E}_*(m)\;\xi(m)\;dm\;,
 \label{eq:edot}
\end{equation}
where $\xi(m)dm$ is the number of stars in the interval $[m,m+dm]$ and $\dot{E}_*(m)$ is the total energy output by a single star of mass $m$. This energy can be calculated from \citep[see][]{bkp08}
\begin{equation}
\log_{10}\dot{E}_* / {\rm erg\;Myr}^{-1} = 50+1.72\left(\log_{10}m/\msun-1.55\right)\;,
\label{eq:starnrg}
\end{equation}
i.e. the contribution through low-mass stars is negligible. Within a gas-removal time-scale, $\tM$, an amount of energy equivalent to $\EOB(\adrei)=\Edot(\adrei)\cdot\tM$ is released by massive stars. The provided energy depends on the shape of the stellar IMF (equation~\ref{eq:edot}) and therefore on the choice of $\adrei$ and $\mmax$. The latter has been found to have a negligible influence on the results and is arbitrarily chosen as $\mmax=120\msun$. In order to evaluate equation~(\ref{eq:edot}), the coefficients $k$ and $a_i$ in equation~(\ref{eq:imf}) are found by normalizing the \emph{canonical} IMF to $\mecl$,
\begin{equation}
 \mecl=\int_{0.08\msun}^{\mmax}m\;\xi_c(m)\;dm\;.
 \label{eq:norm}
\end{equation}
By then changing $\adrei$ individually for the initial condition of each cluster \citep[as constrained by][]{mk10} and calculating equation~(\ref{eq:edot}), a solution for $\adrei$ is obtained for which $\EOB=\Ereq$, such that the energy is sufficient to remove the residual-gas. Using the same $k$ and $a_i$ for all $\adrei$ ensures that the number of low-mass stars, which provide the light after the massive stars extinguish, does not change.

\section{Results} \label{sec:result}
\begin{figure}
\begin{center}
  \includegraphics[width=0.47\textwidth]{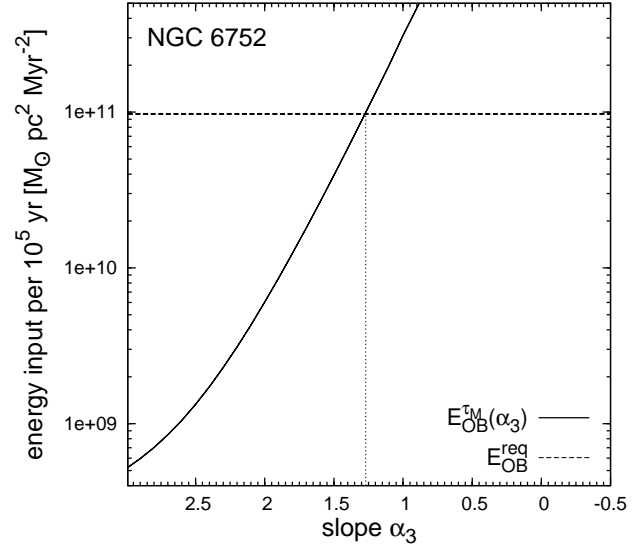}
  \caption{Energy provided by O- and B-type stars within $10^5$~yr as a function of the high-mass IMF index, $\adrei$, for the Galactic GC NGC~6752 (solid curve). The horizontal dashed line indicates the energy, $\Ereq$, required to remove NGC~6752's initial gas mass (equation~\ref{eq:ereq}). For an IMF index $\adrei=1.27$, the energy input from massive stars within $10^5$~yr is sufficient to overcome $\Ereq$.}
  \label{fig:alpha_energy}
\end{center}
\end{figure}
\begin{table*}
\begin{center}
\caption{The birth conditions for the GCs for which the PDMF has been measured from $0.8$ down to $0.3\msun$ by \citet{dmpp07}. Column 1 and 2 list the cluster name and its metallicity. The next five columns give the constraints \citep{mk10} on the inital GC cloud-core mass ($\mcl$), half-mass radius ($\rh$), density within the half-mass radius ($\rhocl$) and lower limit to the star formation efficiency ($\epsilon_{\rm low}$). The last five columns are the assumed residual-gas expulsion time-scale, the here calculated required high-mass IMF slope ($\adrei$) and the corresponding number fraction ($\eta_N$, equation~\ref{eq:etan}) and mass fraction ($\eta_M$, equation~\ref{eq:etam}) of stars more massive than $1\msun$, respectively, and the total-mass fraction lost due to stellar evolution ($f_{\rm sev}$) over one Hubble-time.}
\begin{tabular}{lrrrrrrrrrc}
\hline
Cluster & [Fe/H] & $\mcl$ & $\rh$ & $\rhocl$ & $\epsilon_{\rm low}$ & $\tM$ & $\adrei$ & $\eta_N$ & $\eta_M$ & $f_{\rm sev}$ \\
NGC & & [$10^6\msun$] & [pc] & [$10^6\msun$ pc$^{-3}$] &  & [Myr] &  &  &  &  \\
\hline
104 & -0.76 & 9.40 & 0.49 & 9.54 & 0.25 & 0.1 & 1.34 & 0.25 & 0.94 & 0.84 \\
288 & -1.24 & 1.43 & 1.63 & 0.04 & 0.40 & 0.1 & 2.36 & 0.09 & 0.53 & 0.43 \\
2298 & -1.85 & 1.67 & 0.59 & 0.97 & 0.40 & 0.1 & 1.98 & 0.12 & 0.71 & 0.60 \\
Pal 5 & -1.41 & 0.56 & 7.52 & 2$\times10^{-4}$ & 0.40 & 0.1 & - & - & - & - \\
5139 & -1.62 & 23.10 & 0.87 & 4.19 & 0.25 & 0.1 & 1.26 & 0.28 & 0.96 & 0.85 \\
5272 & -1.57 & 5.48 & 0.47 & 6.30 & 0.33 & 0.1 & 1.54 & 0.19 & 0.90 & 0.79 \\
6121 & -1.20 & 8.78 & 0.42 & 14.15 & 0.25 & 0.1 & 1.32 & 0.26 & 0.95 & 0.84 \\
6218 & -1.48 & 2.72 & 0.64 & 1.24 & 0.33 & 0.1 & 1.81 & 0.15 & 0.79 & 0.68 \\
6254 & -1.52 & 3.39 & 0.42 & 5.46 & 0.25 & 0.1 & 1.56 & 0.19 & 0.89 & 0.78 \\
6341 & -2.28 & 6.73 & 0.23 & 66.03 & 0.15 & 0.1 & 1.11 & 0.34 & 0.98 & 0.87 \\
6352 & -0.70 & 0.78 & 1.08 & 0.07 & 0.40 & 0.1 & 2.43 & 0.09 & 0.50 & 0.41 \\
6397 & -1.95 & 1.76 & 0.17 & 42.76 & 0.15 & 0.1 & 1.37 & 0.24 & 0.94 & 0.83 \\
6496 & -0.64 & 1.33 & 1.18 & 0.10 & 0.25 & 0.1 & 2.10 & 0.11 & 0.65 & 0.55 \\
6656 & -1.64 & 4.52 & 0.42 & 7.28 & 0.20 & 0.1 & 1.43 & 0.22 & 0.93 & 0.82 \\
6712 & -1.01 & 2.17 & 1.18 & 0.16 & 0.50 & 0.1 & 2.20 & 0.10 & 0.60 & 0.50 \\
6752 & -1.56 & 3.68 & 0.24 & 31.78 & 0.15 & 0.1 & 1.27 & 0.27 & 0.96 & 0.85 \\
6809 & -1.81 & 1.99 & 0.85 & 0.39 & 0.25 & 0.1 & 1.89 & 0.13 & 0.76 & 0.65 \\
6838 & -0.73 & 0.30 & 0.47 & 0.34 & 0.50 & 0.1 & 2.60 & 0.08 & 0.44 & 0.35 \\
7078 & -2.16 & 17.30 & 0.20 & 258.13 & 0.10 & 0.1 & 0.76 & 0.56 & 0.99 & 0.89 \\
7099 & -2.12 & 3.30 & 0.23 & 32.38 & 0.15 & 0.1 & 1.29 & 0.27 & 0.95 & 0.85 \\
\hline
\end{tabular}
\label{tab:tab}
\end{center}
\end{table*}
Fig.~\ref{fig:alpha_energy} shows how the energy-input through massive stars changes with the IMF index $\adrei$ using the example of NGC~6752. The steeper the IMF (the lower $\adrei$), the more massive stars are available and the more energy is deposited in the ISM (solid curve). NGC~6752 had to have a top-heavy IMF with $\adrei=1.27$ in order for the OB-star energy input to equal the energy required to remove the gas (dashed line) from its progenitor ($\EOB=\Ereq$).

The results for the estimated $\adrei$-values in the sample of GCs are summarized in Tab.~\ref{tab:tab}. For the respective initial conditions ($\mcl$, $\rh$ and $\epsilon$, also listed in Tab.~\ref{tab:tab}) the results imply that the IMFs above $1\msun$ have been mostly top-heavy ($\adrei\approx1\ldots2.3$, assuming the residual-gas is removed by radiation and winds from stars only). Slightly top-light IMFs are found for NGC~288, 6352 and 6838. A solution for Pal~5 could not be found unless an extremely top-light MF is allowed (i.e. a canonical IMF more than suffices to remove the gas in Pal~5). This is likely connected to Pal~5 being close to dissolution, as already discussed in \citet{mk10}.

\begin{table*}
\begin{center}
\caption{Properties of starburst clusters in the MW and the LMC. From left to right the columns denote the cluster name, its age, stellar mass, $\mecl$, half-mass radius, $\rh$, stellar density, $\rhoecl$, within $\rh$, and range of published high-mass PDMF slopes, $\adrei$, from the references given in the last column.}
\begin{tabular}{rcccccc}
\hline
Name & age/Myr & $\mecl/\msun$ & $\rh/$pc & $\rhoecl/\mpc$ & $\adrei$ & References\\
\hline
Arches & 2-3 & $1.3\times10^4$ & $0.24$ & $1.12\times10^5$ & 1.65-2.3 & 1-6\\
NGC~3606 & 1-3 & $0.7\times10^4$ & $0.2$ & $1.04\times10^5$ & 1.9-2.3& 3,4,7-9\\
Wd1 & 3-4 & $5\times10^4$ & $1$ & $5.96\times10^3$ & 1.6-2.3 & 10\\
R136 & 2-3 & $10^5$ & $1.1$ & $8.97\times10^3$ & 2.2-2.6 & 11-13\\
\hline
\multicolumn{7}{l}{$^1$\citet{Figer1999}, $^2$\citet{Stolte2002}, $^3$\citet{Stolte2005}, $^4$\citet{Stolte2006}, $^5$\citet{Kim2006}}\\
\multicolumn{7}{l}{$^6$\citet{Espinoza2009}, $^7$ \citet{Nuerberger2002}, $^8$\citet{Sung2004}, $^9$\citet{Harayama2008}}\\
\multicolumn{7}{l}{$^{10}$\citet{Brandner2008}, $^{11}$\citet{Andersen2009}, $^{12}$\citet{Brandl1996}, $^{13}$\citet{Massey1998}}
\end{tabular}
\label{tab:clust}
\end{center}
\end{table*}
To put these results into context, they are compared to independent theoretical and observational evidence for the existence of top-heavy IMFs in UCDs. Also, a comparison for different IMF observations in present-day Local Group starburst clusters (see Sec.~\ref{sec:intro}) is performed.

For the starburst clusters their respective birth cloud-core masses and densities are needed for this comparison. These are here estimated using their present-day masses, half-mass radii and densities as compiled in Table~\ref{tab:clust} and assuming that the clusters have formed with $\epsilon=1/3$. Given the youth of these clusters, no significant mass loss apart from gas-blow out is assumed to have occurred. Their cloud-core densities at birth, $\rhocl$, were calculated similarly and assuming that their sizes did not yet change strongly during gas expulsion. The initial conditions for UCDs are readily available from the \citet{dab10} simulations.

\begin{figure}
\begin{center}
 \includegraphics[width=0.47\textwidth]{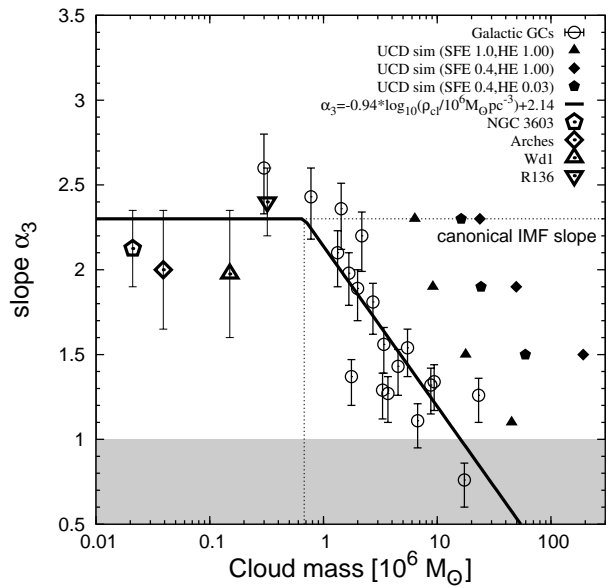}
 \caption{The high-mass IMF slope, $\adrei$, for GCs (open circles with errorbars) decreases with increasing pre-GC cloud-core mass, $\mcl$ (stars+gas). The solid line is a least-squares fit to the GC data and the corresponding equation is indicated in the legend. The IMF becomes canonical ($\adrei=2.3$, horizontal dotted line) below $6.8\times10^5\msun$ (vertical dotted line). The range of quoted MF slope values found in the literature for the massive, young clusters NGC 3603, Arches, Wd1 and R136 are indicated as the open pentagon, diamond, up- and downward triangle, respectively, with errorbars, being derived from the range of published $\adrei$ measurements. The overall trend is that more massive GCs form relatively more massive-stars (flatter IMFs). The filled symbols correspond to the initial conditions of UCD simulations that lead, after primarily gas expulsion and stellar mass loss driven evolution, to objects that resemble the properties of observed UCDs today. Different symbols correspond to different input parameters in the UCD models \citep[star formation efficiency SFE and heating efficiency HE, see][]{dab10}. The UCD data form a separate group that runs roughly parallel to the GC data. Survival of GCs with $\adrei<1$ (grey-shading) is questionable (Sec.~\ref{sec:discuss}).}
 \label{fig:alpha_mass}
\end{center}
\end{figure}
\begin{figure}
\begin{center}
 \includegraphics[width=0.47\textwidth]{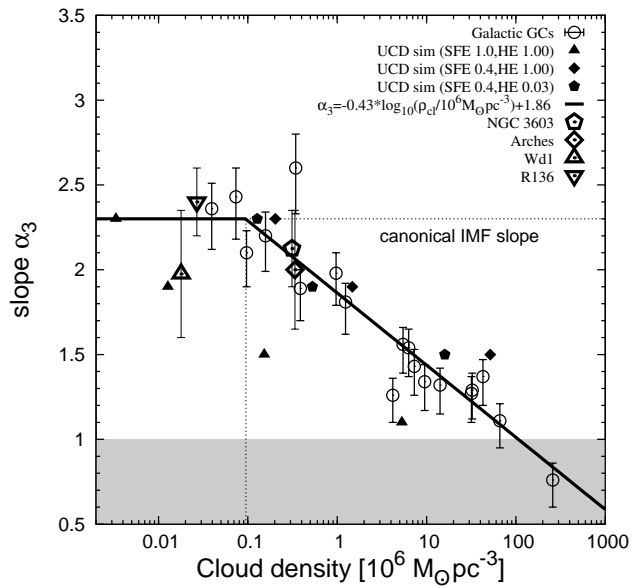}
 \caption{The high-mass IMF index, $\adrei$, versus the cloud-core density, $\rhocl$ (stars+gas). Symbols and lines as in Fig.~\ref{fig:alpha_mass}. The gap between the GCs and UCDs seen in Fig.~\ref{fig:alpha_mass} vanishes and the IMF indices, $\adrei$, now form a single sequence as a function of $\rhocl$: The IMF index at the high-mass end decreases (the IMF becomes flatter) with increasing density of the objects' progenitor cloud-core. It is canonical ($\adrei=2.3$) below $9.5\times10^4\msun$ pc$^{-3}$.}
 \label{fig:alpha_density}
\end{center}
\end{figure}
Fig.~\ref{fig:alpha_mass} shows the dependence of the high-mass IMF slope as a function of the pre-gas expulsion cloud-core mass. While the $\adrei(\mcl)$-trend for UCD models and the constraints for GCs+starburst clusters are similar, both kinds of object appear to form separate sequences running parallel to each other. However, when describing $\adrei$ as a function of the birth cloud-core mass-density, $\rhocl$, i.e. incorporating information of the objects' sizes, the independently obtained data agree excellently with each other (Fig.~\ref{fig:alpha_density}).

\emph{Figs.~\ref{fig:alpha_mass} and \ref{fig:alpha_density} thus suggest that the high-mass IMF was more top-heavy (flatter) in more massive and denser environments.} The expectation would indeed be that denser systems form more massive stars with respect to the canonical IMF, if a channel for massive-star formation is the coagulation of proto-stellar cores since the collision probability is higher in denser systems, where the stars may also have larger accretion rates \citep[Sec.~\ref{sec:intro}]{ml96} and/or the \citet{Pap10} cosmic ray heating is active in star bursts.

\begin{figure}
\begin{center}
 \includegraphics[width=0.47\textwidth]{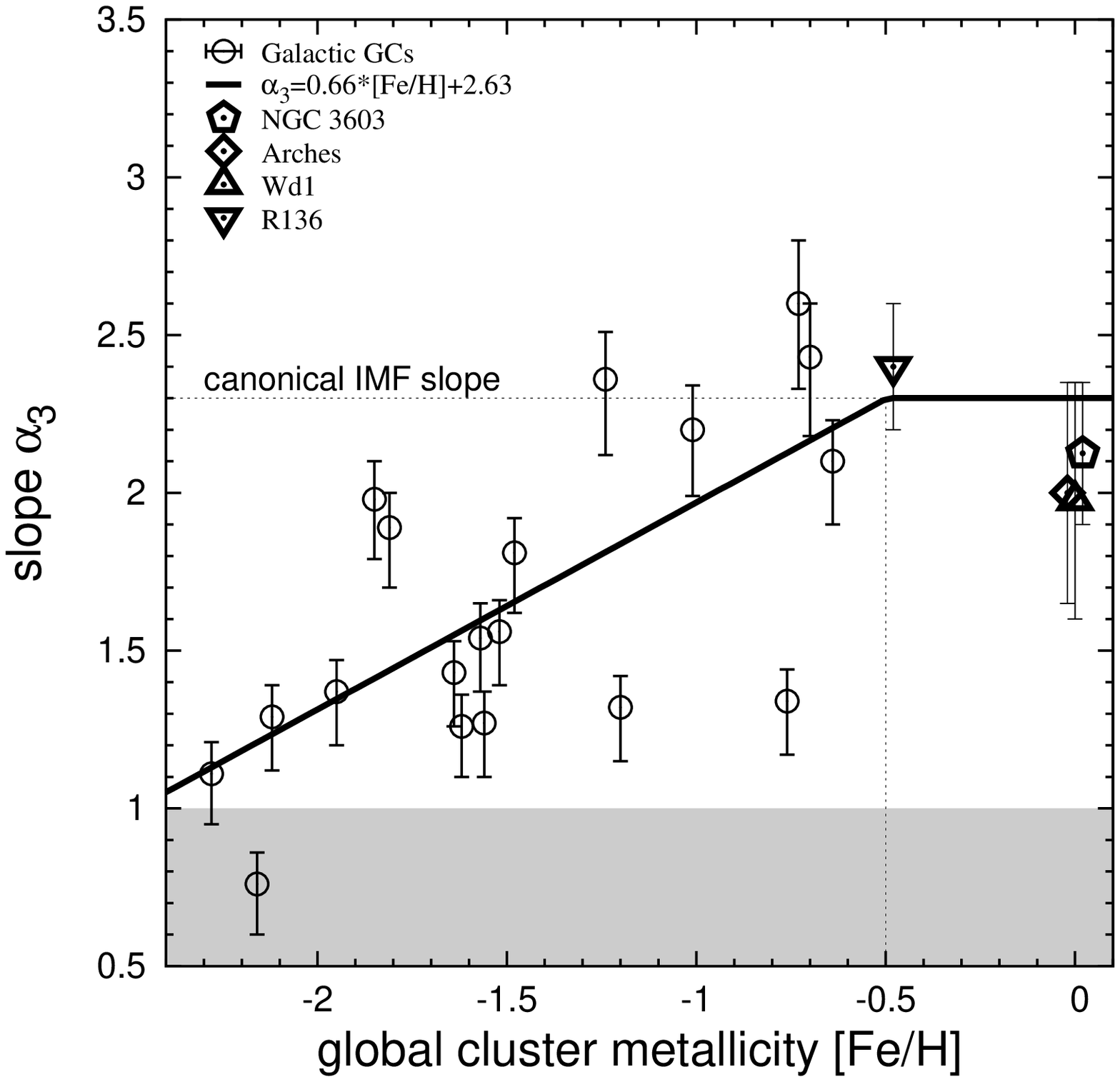}
  \caption{The high-mass IMF index, $\adrei$, as a function of the global cluster metallicity, [Fe/H]. Symbols and lines as in Figs.~\ref{fig:alpha_mass} and~\ref{fig:alpha_density}. NGC 3603, Arches and Wd1 have about solar metallicity. For R136 [Fe/H]$\approx-0.48$ is adopted, a typical value for the LMC. The data suggest the IMF to become less top-heavy with increasing cluster metallicity, having the canonical value ($\adrei=2.3$) above ${\rm [Fe/H]}=-0.5$.}
 \label{fig:alpha_metal}
\end{center}
\end{figure}
Finally, Fig.~\ref{fig:alpha_metal} depicts  $\adrei$ as a function of the present-day global metallicity for the GC and Local Group starburst cluster data. \emph{Metal-poorer environments appear to form flatter IMFs.} Again, this reflects the theoretical expectation: In systems of low metallicity the Jeans mass is larger favouring the formation of more high-mass stars \citep{AdFat96,l98}. A straight-forward least-squares fit to the GC data suggests the IMF to become canonical above $[Fe/H]=-0.5$.

\begin{table}
\begin{center}
\caption{Coefficients (equation~\ref{eq:fit}) for least-squares fits to the GC data (Figs.~\ref{fig:alpha_mass},~\ref{fig:alpha_density} and~\ref{fig:alpha_metal}). Additionally, coefficients for the dependence of $\adrei$ on the stellar mass ($\mecl=\epsilon\times\mcl$) forming from the cloud-cores are given. $\mecl$ and $\mcl$ are in units of $10^6\msun$ and $\rhocl$ is in units of $10^6\msun$~pc$^{-3}$.}
\begin{tabular}{rcccc}
\hline
$\lambda$ & $p_{\lambda}$ & $q_{\lambda}$ & $\lambda_{\rm lim}$ & Fig. \\
\hline
$\log_{10}\mcl$ & -0.94 & 2.14 & $>6.8\times10^5\msun$ & \ref{fig:alpha_mass} \\
$\log_{10}\mecl$ & -0.77 & 1.59 & $>2.7\times10^5\msun$ & - \\
$\log_{10}\rhocl$ & -0.43 & 1.86 & $>9.5\times10^4\msun$~pc$^{-3}$ & \ref{fig:alpha_density} \\
$[Fe/H]$ & 0.66 & 2.63 & $<-0.5$ & \ref{fig:alpha_metal} \\
\hline
\end{tabular}
\label{tab:coeff}
\end{center}
\end{table}
\emph{Thus, evidence for a systematic variation of the high-mass IMF in dependence of the birth environment emerges.} In the following, each dependence of the suggested top-heaviness of Galactic GCs on their progenitor cloud-core properties is described by linear relations. Writing
\begin{equation}
 \adrei\left(\lambda\right)=\left\{\begin{array}{rl}
                p_{\lambda}\times\lambda+q_{\lambda}, & \lambda\gtrless\lambda_{\rm lim}\;, \\
                2.3, & {\rm otherwise}\;,
               \end{array}
 \right.
 \label{eq:fit}
\end{equation}
$\lambda$ is either $\log_{10}(\mcl/10^6\msun)$ (Fig.~\ref{fig:alpha_mass}), $\log_{10}(\rhocl/10^6\msun)$~pc$^{-3}$ (Fig.~\ref{fig:alpha_density}) or $[Fe/H]$ (Fig.~\ref{fig:alpha_metal}) with the corresponding coefficients, $p_{\lambda}$ and $q_{\lambda}$. The parameter $\lambda_{\rm lim}$ is the limiting value above or below which the IMF is top-heavy. The parameters are shown in Tab.~\ref{tab:coeff}, with the appropriate inequality sign ($<$ or $>$) for $\lambda_{\rm lim}$ being reported.

\begin{figure}
\begin{center}
 \includegraphics[width=0.47\textwidth]{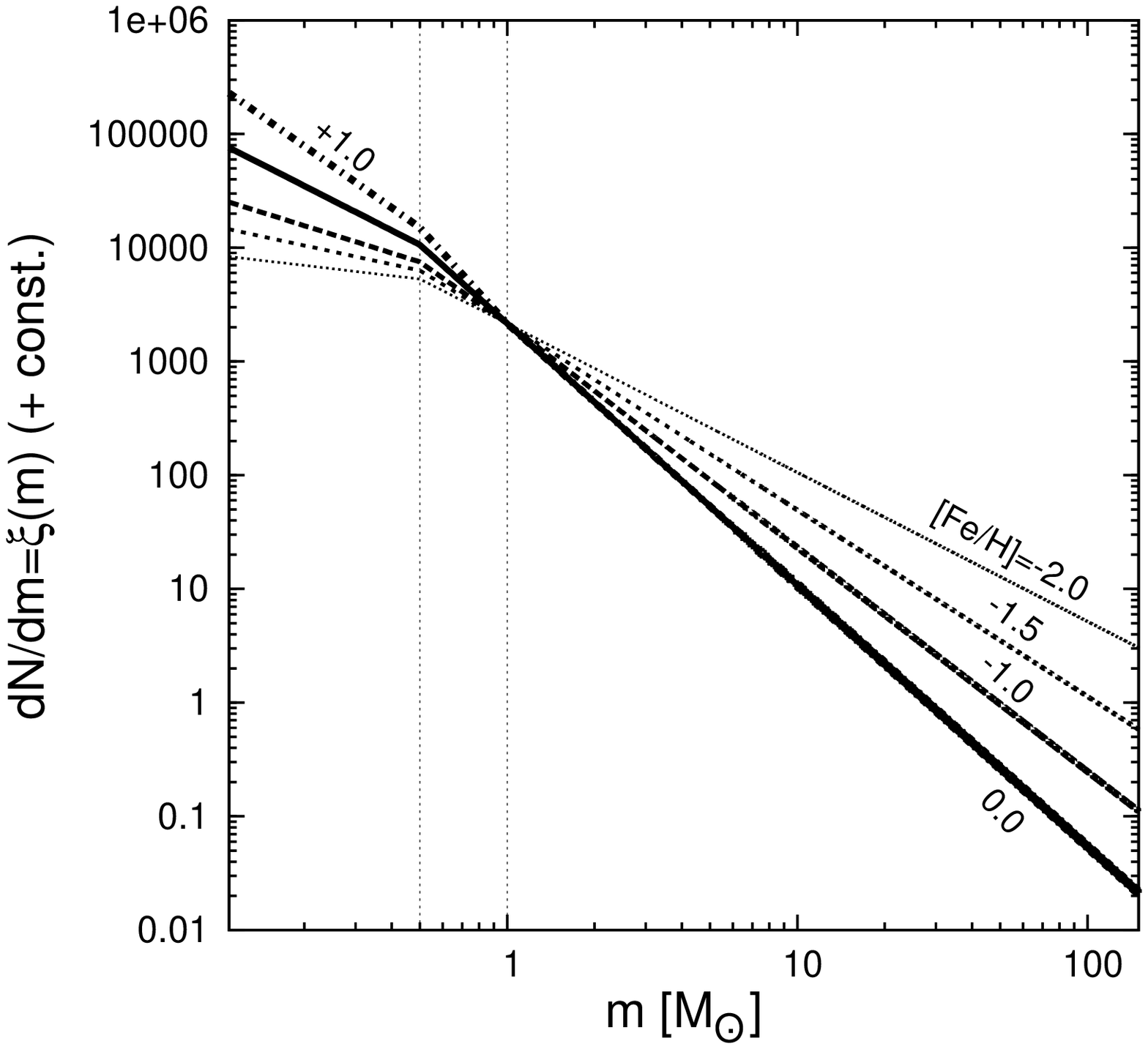}
  \caption{Suggested shape of the stellar IMF for different metallicities, [Fe/H] (not taking into account the density dependence of the IMF). The IMFs are scaled such that their values agree at $m=1\msun$. Above $1\msun$ the IMF slope is determined by the present work (Fig.~\ref{fig:alpha_metal}, equation~\ref{eq:fit}). Below $1\msun$ the parametrisation is by \citet[equation~\ref{eq:lowvar}]{k01}, whose results suggest tentative evidence that more metal-rich environments produce relatively more low-mass stars. Note that only the metallicity dependence is shown, but not the dependence on mass (Fig.~\ref{fig:alpha_mass}) or density (Fig.~\ref{fig:alpha_density}).}
 \label{fig:imf_metal}
\end{center}
\end{figure}
\begin{table}
\begin{center}
\caption{IMF indices, $\alpha_i$, in equation~(\ref{eq:imf}) in dependence of the metallicity, [Fe/H]. The indices $\alpha_{1/2}$ are calculated from equation~(\ref{eq:lowvar}), and $\adrei$ is obtained using equation~(\ref{eq:fit}) with the parameters in Tab.~\ref{tab:coeff}. For [Fe/H]$=0.0$ the IMF is canonical everywhere. The IMF metallicity-variation is depicted in Fig.~\ref{fig:imf_metal}.}
\begin{tabular}{rccc}
\hline
[Fe/H] & $\alpha_1$ & $\alpha_2$ & $\adrei$ \\
\hline
-2.0 & 0.30 & 1.30 & 1.31 \\
-1.5 & 0.55 & 1.55 & 1.64 \\
-1.0 & 0.80 & 1.80 & 1.97 \\
 0.0 & 1.30 & 2.30 & 2.30 \\
+0.5 & 1.55 & 2.55 & 2.30 \\
+1.0 & 1.80 & 2.80 & 2.30 \\
\hline
\end{tabular}
\label{tab:indices}
\end{center}
\end{table}
The variation of the stellar IMF based on the metallicity is summarized in Fig.~\ref{fig:imf_metal}. Above a stellar mass of $1\msun$ the IMF flattens with decreasing metallicity (Fig.~\ref{fig:alpha_metal}, equation~\ref{eq:fit}). By considering young and intermediate-age open clusters \citet{k01} presented evidence that the IMF might also be dependent on [Fe/H] below $1\msun$ in the sense that a flatter low-mass IMF occurs for lower metallicities. His parametrisation for low-mass IMF variation is adopted here,
\begin{equation}
 \alpha_{1/2}=\alpha_{1/2,\rm c}+\Delta\alpha{\rm [Fe/H]}\;,
 \label{eq:lowvar}
\end{equation}
where $\Delta\alpha\approx0.5$ and $\alpha_{1/2,\rm c}$ are the respective slopes of the canonical IMF (Sec.~\ref{sec:tM}). The variation described by equation~\ref{eq:lowvar} is incorporated in Fig.~\ref{fig:imf_metal}. Note that this parametrisation has been suggested for metallicities $\gtrsim-0.5$. For the lower metallicities needed here the calculated low-mass IMF slopes are thus extrapolated values. Furthermore this variation is not incorporated in the \citet{mk10} models and cannot be used to explain the observed low-mass star depletion in the observed GCs: Using equation~\ref{eq:lowvar} the IMF becomes steeper with increasing metallicity, contrary to observations of the PDMFs (Sec.~\ref{sec:intro}). The resulting IMF indices, $\alpha_i$, are reported in Tab.~\ref{tab:indices}.

\section{A fundamental plane for $\adrei$, $\log\rhocl$ and [Fe/H]} \label{sec:plane}
In this section a formula, $\adrei(\log\rhocl,{\rm [Fe/H]})$, is derived in order to describe the dependence of $\adrei$ on density and metallicity for GCs simultaneously. The idea is to find a \emph{fundamental plane}, $p$, in ($\adrei,\log\rhocl,{\rm [Fe/H]}$)-space which minimizes the scatter of $\adrei$-values around $p$.

\begin{figure}
\begin{center}
 \includegraphics[width=0.47\textwidth]{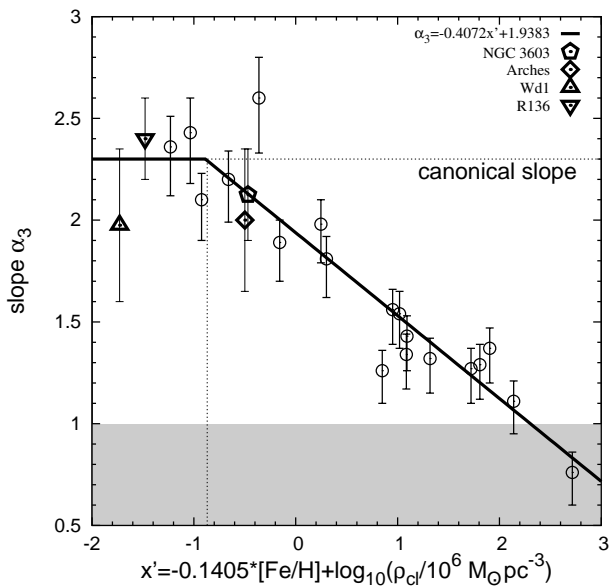}
  \caption{Edge-on view of the fundamental plane, $p$, (equation~\ref{eq:plane}, solid regression line) for the IMF index, $\adrei$, cluster density, $\log_{10}\rhocl$, and metallicity, [Fe/H]. The abscissa is the new axis, $x'$, in which the scatter around the solid regression line through the data projected onto the $\adrei$--$x'$ plane is minimal (equation~\ref{eq:scat}).}
 \label{fig:plane}
\end{center}
\end{figure}
The method used to find $p$ follows the same basic principle as \citet{Kroupa2010} used to determine the orientation of the disc of Milky Way satellites. The coordinate system ($\adrei,\log\rhocl,{\rm [Fe/H]}$) is rotated stepwise by $\Delta\vartheta=1^\circ$ around the $\adrei$-axis. In each step the original data are projected onto the new $\adrei$--$x'(\vartheta)$ plane, where
\begin{equation}
 x'(\vartheta)=\cos\vartheta\times{\rm [Fe/H]}+\sin\vartheta\times\log_{10}(\rhocl/10^6\;\mpc),
 \label{eq:scat}
\end{equation}
and a linear regression in this plane is performed. The rotation angle $\vartheta$ in which the root mean squared-deviation of the projected data from the regression line along the $\adrei$-axis is minimal is the angle with the least scatter. For the data in Tab.~\ref{tab:tab} this is the case for $\vartheta=98^\circ$, i.e. $x'$ lies close to the original $\adrei$--$\rhocl$ plane.

Fig.~\ref{fig:plane} depicts $x'$ (actually $x'/\sin(98^\circ)$) versus the IMF index $\adrei$. The regression line through the data in this representation is the fundamental plane, $p$, seen edge-on (i.e. the projection of $p$ onto the $\adrei$--$x'$ plane). The fundamental plane equation thus reads
\begin{eqnarray}
 p&\equiv&\adrei(x')\nonumber\\
  &=&\left\{\begin{array}{rl}
      -0.4072\times x'+1.9383 & x'\geq0.87\\
      2.3 & {\rm otherwise}
     \end{array}\right.
\label{eq:plane}
\end{eqnarray}
and making use of equation~(\ref{eq:scat}) for $\vartheta=98^\circ$ and $x'\geq0.87$,
\begin{eqnarray}
  &&\adrei(\log\rhocl,{\rm [Fe/H]})=0.0572\times{\rm [Fe/H]}\nonumber \\
  &&-0.4072\times\log_{10}(\rhocl/10^6\;\mpc)+1.9383\;.
  \label{eq:plane2}
\end{eqnarray}
As evident from its larger coefficient in equation~(\ref{eq:plane2}), density has a stronger influence on $\adrei$ than metallicity. This is expected given the significantly lower scatter in Fig.~\ref{fig:alpha_density} in comparison with Fig.~\ref{fig:alpha_metal}. The fundamental plane has only a slightly smaller scatter in $\adrei$ than the $\adrei$--$\log\rhocl$ plane. The summed squared-residuals are~$0.56$ and~$0.57$, respectively.

The stronger dependence on $\log\rhocl$ can be understood as follows: A dense cluster forming from gas with solar metallicity will still produce a top-heavy IMF, although the [Fe/H] dependence alone (Fig.~\ref{fig:alpha_metal}) suggests the canonical slope. Vice versa, a more extended cluster with low-[Fe/H] will not be as top-heavy as expected from Fig.~\ref{fig:alpha_metal}.

The procedure used here is similar, but not equivalent, to a \emph{principal component analysis}.

\section{Survival of GCs with top-heavy initial mass functions and caveats} \label{sec:discuss}
The combined information from this work, from the independent analysis of the top-heaviness in UCDs and the (rare and weak) observational evidence of top-heavy IMFs in starburst clusters, which are remarkably consistent with each other, suggests that some presently old dense stellar systems might have formed stars which were selected from a top-heavy IMF. The flatness of the high-mass end of the IMF reflects the metal-content and the density of the environment in which the respective objects were born.

The question may be raised whether embedded clusters with extremely top-heavy IMFs (in particular $\adrei\lesssim1$) would survive stellar evolution and, if they do, whether they are long-lived, because a significant amount of mass would be lost from massive evolving stars, triggering cluster expansion. Actually, for $\adrei\lesssim1$ (grey-shaded area in Figs.~\ref{fig:alpha_mass},~\ref{fig:alpha_density},~\ref{fig:alpha_metal} and~\ref{fig:plane}) almost the total mass of the cluster is contained in stars with $m>1\msun$ and $\approx90$ per cent of the total mass will be lost due to stellar evolution. In the best case this mass is lost adiabatically, i.e. slowly with respect to the crossing-time of the cluster. The cluster then expands by a factor of $10$ (equation~\ref{eq:hills}) through stellar evolution alone. The expansion due to stellar evolution becomes stronger if the cluster is mass-segregated \citep{Vesp09}, as the cluster models in \citet{mk10} are. In Tab.~\ref{tab:tab} the number- and mass-fractions of stars more massive than $1\msun$,
\begin{equation}
 \eta_N=\frac{\int_{1\msun}^{\mmax}\xi_c(m)dm}{\int_{0.08\msun}^{\mmax}\xi_c(m)dm}\;,
 \label{eq:etan}
\end{equation}
and
\begin{equation}
 \eta_M=\frac{\int_{1\msun}^{\mmax}m\;\xi_c(m)dm}{\int_{0.08\msun}^{\mmax}m\;\xi_c(m)dm}\;,
 \label{eq:etam}
\end{equation}
respectively, as well as the total mass lost due to stellar evolution \citep[following][]{dab10} for the here estimated $\adrei$-values are reported.

L\"ughausen et al. (in preparation) investigate, by means of $N$-body computations, the survival of star clusters with a top-heavy IMF having stellar masses between 5000 and 25000 $\msun$ and half-mass radii between $\approx0.4$ and $1$~pc. They find that their models, having average stellar densities within the half-mass radius up to $\approx2\times10^4\msun$~pc$^{-3}$, with an IMF index $\alpha_3\leq1.0$ immediately dissolve through strong stellar evolution driven mass loss and an accelerated dynamics (for a fixed cluster mass the relaxation-time becomes shorter with increasing top-heaviness due to the smaller number of stars). Slightly less top-heavy models are not very long-lived. The birth stellar densities for the Galactic GCs (Tab.~\ref{tab:tab}) are however two to four orders of magnitudes larger than the density-range covered by L\"ughausen et al. (in preparation). In particular survival is likely to become possible when the crossing time, $\tcr$, becomes shorter than the stellar evolution time-scale $m/\dot{m}$ for massive stars, where $\dot{m}$ is the mass-loss rate of a star of mass $m$. Therefore survival of such dense objects with extremely top-heavy IMFs remains to be checked numerically.

The constraints on the top-heaviness were derived assuming the energy for gas removal is provided by stellar winds and radiation only and all energy radiated by the stars is completely absorbed by the gas. This is likely to be an oversimplification \citep{dale05}. If the initially most massive clusters are able to retain the residual-gas in their deep potentials until supernovae occur, their energy input might be sufficient to remove the gas in $\approx10^5$~yr. Additionaly, the metal content might influence the gas throw-out through radiation$\leftrightarrow$gas/dust coupling leading to different energy absorbtion efficiencies \citep{mk10}. Both effects need to be considered in future work.

The parametrisation (equations~\ref{eq:fit} and~\ref{eq:plane}) should therefore not be seen as an established dependence on the cluster initial parameters. Instead it is suggestive of what IMF variation the present observational data suggest.

\section{Conclusions} \label{sec:sum}
This work has shown that in oder to remove the residual-gas within $\approx10^5$~yr after star formation from progenitors of some Galactic GCs \citep{mk10} starting with a universal low-mass IMF, top-heavy IMFs need to be invoked. The IMF index/slope $\adrei$ in equation~(\ref{eq:imf}) takes values between $1$ (very top-heavy) and $2.3$ (the canonical Salpeter/Massey-value). Although long-term survival for some GCs with strongly top-heavy IMFs might be at threat through stellar evolution alone, self-consistent modelling of very dense GCs with a top-heavy IMF is needed to investigate their survival chances. The index $\adrei$ anti-correlates roughly $\log$-linearly with the pre-gas expulsion cloud-core mass and density (Figs.~\ref{fig:alpha_mass} and~\ref{fig:alpha_density}) and correlates linearly with the GC metallicity (Fig.~\ref{fig:alpha_metal}).

Combining the density and metallicity dependence of $\adrei$, a fundamental plane has been found which describes the variation of the IMF with both parameters simultaneously. The fundamental plane suggests that $\adrei$ varies more strongly with density than with metallicity.

Large mass-to-light ratios in UCDs indicate this class of objects to have formed with a top-heavy IMF as well \citep{dab08}. The data from \citet{dab10} on probable initial conditions of observed UCDs form a separate group of objects in the $\adrei-$cloud-core mass diagram but run roughly parallelly to the GC data (Fig.~\ref{fig:alpha_mass}). The disappearance of this gap when $\adrei$ is depicted against the cloud-core density (Fig.~\ref{fig:alpha_density}) might suggest that UCDs form from mergers of GCs in cluster complexes \citep{Fel05,ass11,bruens11}, which moves merged GCs of the same top-heaviness to higher UCD masses in Fig.~\ref{fig:alpha_mass}.

The decreasing trend of $\adrei$ with increasing density is qualitatively understood since denser systems are expected to form more massive stars with respect to the canonical IMF, if a channel for massive-star formation is accretion-driven coagulation of proto-stellar cores due to a higher collision probability \citep{ml96}. And in systems of low metallicity the Jeans mass is expected to be larger favouring the formation of more high-mass stars \citep{AdFat96,l98}.

Are these dependencies compatible with observations of young clusters in the Local Group? For the MW and for the LMC the answer is positive (Figs.~\ref{fig:alpha_mass}, \ref{fig:alpha_density}, \ref{fig:alpha_metal}). Although the known Local Group starburst clusters lie just in the density-regime where the IMF starts to become top-heavy ($\rhocl\gtrsim9.5\times10^4\mpc$) the observations are consistent with the results. For the densities of these clusters, stellar dynamical biases would probably hide any true present top-heavy IMF due to mass segregation \citep{pz07}. \citet{m03} shows homogeneously estimated mass function slopes for OB associations in the MW and LMC finding all of them to be consistent with the Salpeter slope ($\alpha=2.35$). For the LMC clusters ([Fe/H]$\approx-0.4$) and solar metallicity MW clusters this is consistent with the results in this work (Fig.~\ref{fig:alpha_metal}), given the star-forming densities in the Local Group are $<10^5\mpc$.

A similar, but slightly different parametrisation of $\adrei$ with the stellar cluster mass, $\mecl$, (Tab.~\ref{tab:coeff}) has been made use of in \citet{wkp11} to include the evidence for a systematically varying IMF into the framework of the theory of the integrated galactic IMF \citep[IGIMF, i.e. the stellar IMF of whole galaxies,][]{KroupWeid03,WeidKroup05}. Under conditions when the galaxy-wide star formation rate SFR$>27\msun$~yr$^{-1}$ star burst clusters with masses $\mecl>10^6\msun$ form \citep{wkl04}. The IGIMF, being the sum of all IMFs in all young star clusters, then also becomes top-heavy\footnote{Top-heavy here also means relative to the canonical IMF.}. Their models lead to reasonable agreement with cosmological observations. The results obtained here are explicitely incorportated into the IGIMF framework in \citet{KroupaRev2012}, where, for example, it is shown that the IGIMF of the Galactic bulge agrees with the chemical abundance constraints. In which way the chemical evolution of galaxies over cosmological time would be affected by a top-heavy IGIMF is subject to further studies.

To conclude, varying the low-mass IMF in order to explain the observed PDMFs (and, in turn, to avoid top-heavy IMFs) in GCs does not appear to be a feasible solution. GCs formed with the canonical IMF over all stellar masses and evolving secularly over a Hubble time cannot lead to the observed concentration--PDMF correlation, unless they are born mass-segregated and filling their tidal radii. But this begs the question how such distended mass-segregated clusters can form in the first place.

Instead, initially concentrated, dense $\lesssim1$~pc large mass-segregated massive progenitors of present-day GCs uncover evidence for a systematically varying high-mass IMF with density and metallicity, in remarkably good agreement with the independently obtained evidence for top-heavy IMFs in UCDs by \citet{dab10}. The suggested trend is qualitatively in agreement with the expectation that higher density and lower metallicity environments should form more top-heavy IMFs (see Sec.~\ref{sec:intro}). Observations are consistent with the here derived trend but mass function measurements for \emph{initially} denser clusters than NGC~3603 and the Arches cluster are likely needed to conclusively test the here suggested behaviour for the high-mass IMF.
\\\\\textbf{Acknowledgments}
MM was supported for this research through a stipend from the International Max Planck Research School (IMPRS) for Astronomy and Astrophysics at the Universities of Bonn and Cologne. JD thanks the DFG for support through grant KR1635/13. MSP acknowledges support through DFG research grant KR 1635/18-2 in the frame of the DFG Priority Programme 1177, Witnesses of Cosmic History: Formation and evolution of galaxies, black holes, and their environment and through the Bonn-Cologne Graduate School of Physics and Astronomy.

\bibliographystyle{mn2e}
\bibliography{biblio}
\makeatletter   \renewcommand{\@biblabel}[1]{[#1]}   \makeatother

\label{lastpage}

\end{document}